\documentclass[preprint,onecolumn,11pt,showpacs,showmcs,showkeys,amsmath,amssymb,floatfix]{revtex4}

\newcommand{\begg}{\begin{gather}}
\newcommand{\beq}{\begin{equation}}

\newcommand{\eegg}{\end{gather}}
\newcommand{\eeq}{\end{equation}}

\usepackage{amssymb}
\usepackage{bm}
\usepackage{color}
\usepackage{epsfig}
\usepackage{dcolumn}
\usepackage{graphicx}
\usepackage{hyperref}
\usepackage{natbib}
\usepackage[normalem]{ulem}
\usepackage{txfonts}

\begin{document}


\title{
Dispersion by pulsars, magnetars, fast radio bursts and massive electromagnetism at very low radio frequencies}


\author{
Mark J. Bentum\textsuperscript{a,b}\footnote{m.j.bentum@utwente.nl},
Luca Bonetti\textsuperscript{c,d}\footnote{luca.bonetti@cnrs-orleans.fr}, 
Alessandro D.A.M. Spallicci\textsuperscript{c,d}\footnote{Corresponding author, spallicci@cnrs-orleans.fr, http://lpc2e.cnrs-orleans.fr/$\sim$spallicci/}
}

\address{\textsuperscript{a}Universiteit van Twente\\
\mbox{Faculteit van  Elektrotechniek, Wiskunde en Informatica, 
Telecommunication Engineering}\\
Postbus 217, 7500 AE Enschede, Nederland\\
\textsuperscript{b}ASTRON\\ 
Postbus 2, 7900 AA Dwingeloo, Nederland \\       
\textsuperscript{c}Universit\'e d'Orl\'eans\\
Observatoire des Sciences de l'Univers en r\'egion Centre\\
\mbox{1A rue de la F\'{e}rollerie, 45071 Orl\'{e}ans, France}\\
\mbox{Collegium Sciences et Techniques, P\^ole de Physique }\\ 
Rue de Chartres, 45100  Orl\'{e}ans, France\\
\textsuperscript{d}Centre Nationale de la Recherche Scientifique\\ 
Laboratoire de Physique et Chimie de l'Environnement et de l'Espace\\ 
\mbox {3A Av. de la Recherche Scientifique, 45071 Orl\'eans, France}}

\begin{abstract}
Our understanding of the universe relies mostly on electromagnetism. As photons are the messengers, fundamental physics is concerned in testing their properties. Photon mass upper limits have been earlier set through pulsar observations, but new investigations are offered by 
the excess of dispersion measure (DM) sometimes observed with pulsar and magnetar data at low frequencies, or with the fast radio bursts (FRBs), of yet unknown origin. Arguments for the excess of DM do not reach a consensus, but are not mutually exclusive. Thus, we remind that for massive electromagnetism, dispersion goes as the inverse of the frequency squared. Thereby, new avenues are offered also by the recently operating ground observatories in 10-80 MHz domain and by the proposed Orbiting Low Frequency Antennas for Radio astronomy (OLFAR ). The latter acts as a large aperture dish by employing a swarm of nano-satellites observing the sky for the first time in the 0.1 - 15 MHz spectrum. 
The swarm must be deployed sufficiently away from the ionosphere to avoid distortions especially during the solar maxima, terrestrial
interference and offer stable conditions for calibration during observations. 
\end{abstract}

\date{17 October 2016}

\maketitle
 
\section{Motivations}

For many years, many of us have worked, see {\it e.g.}, \cite{spalliccietal2005}, for the opening of one of the gravitational wave windows on the universe, that has finally occurred \cite{abbottetal2016}, while for other windows we still have to wait \cite{spallicci2013}. 
Nevertheless, we can genuinely state that even when gravitational wave information will be exploited by ground or space laser interferometers, or by the pulsar timing array, our understanding of the large scale universe will be largely (exception made for neutrinos and cosmic rays) based on electromagnetic observations of the four interactions that rule the physical world. As photons are the messengers, fundamental physics has the concern of testing the foundations of electromagnetism, while astrophysics the task of interpreting the universe accordingly. 

{ Furthermore, while alternative theories to general relativity, including those based on massive gravitons, are also conceived for solving the questions raised by the dark universe or to couple gravity with the other interactions, less effort is deployed for studying alternative electromagnetism. But electromagnetism  at large scales may differ from the Maxwellian conception of the nineteenth century and thereby contribute to solve some of the riddles in contemporary physics and cosmology}. Since neutrinos have been declared massive, while waiting for the elusive graviton, and leaving aside gluons that are not observed as free particles, the photon is the sole particle of the standard model to be massless. 

Finally, massive photons would manifest themselves through delays of low frequency electromagnetic signals, being the sub-MHz part of the spectrum yet unexplored \cite{lacki2010}.  Thus, this work addresses the potential of the low frequency region to test the foundations of physics.  

We shall use SI units throughout the paper. 

\section{Massive electromagnetism}

Non-Maxwellian theories can be grouped into two main classes: non-linear massless theories and linear massive photon theories. The former were initiated by Born and Infeld \cite{boin34} who remove the divergence of the electron self-energy by introducing an upper bound of the electric field at the origin, and by Heisenberg and Euler \cite{heeu36} who
describe the non-linear dynamics of electromagnetic fields in vacuum and predict the rates of quantum electrodynamics light interaction processes. The analysis of the propagation of photons in vacuum within theories of non-linear electrodynamics is carried by supposing the photon to propagate in an effective metric. The latter represents the underlying dynamics that modifies the structure of the geometry in such a way that the geometry on which the photons move is no longer the background geometry. 
In the linear Maxwell theory,  there is no influence of a background field on the propagation of light. In the non-linear theories, however,
the phase velocity of light depends on the strength of the background field and on the propagation direction
relative to the background field.
Photon-photon scattering \cite{breitwheeler1934,halpern1934,euler1936,heeu36,achieser1937,karplusneuman1950,karplusneuman1951} has been tested \cite{burkeetal1997}, while for other experiments aiming to verify other features of non-linear theories see \cite{scpela15} and references therein. 

The latter class assume or predict the photon being massive. Massive photons have been evoked in different realms: dark matter and dark energy, inflation, charge conservation, magnetic monopoles, Higgs boson, non-cosmological redshifts; in applied physics, superconductors and "light shining through walls" experiments. The mass can be considered effective, when supposed depending on given parameters.

The concept of a massive photon was pursued by de Broglie \cite{db22,db23,db40}, also thanks to the work of his doctorate student Proca, originally aimed at the description of electrons and positrons \cite{pr36a,pr36b,pr36c,pr36d,pr37,pr38}. 

How much the theory of relativity would be affected by the massive photon assumption, is not straightforward to assess: due partly to the variety of the theories, and partly to the removal of our ordinary landmarks and the rising of interwoven implications. For instance, the de Broglie-Proca, henceforth dBP, equations follow special relativity laws for reference frames moving at constant velocities (they obey Lorentz-Poincar\'e's transformations and hence the Lagrangian can be written covariantly); instead, a change of potential implies a change in the field (the equations are not Lorenz gauge invariant). 

Massive electromagnetic theories, possibly gauge invariant, have been later proposed by several authors, see  \cite{bopp1940,podolski1942,podolskikikuchi1944,podolskyschwed1948,st57,chernsimons1974} to mention the earliest contributors. 
Also the works \cite{lande1941,landethomas1941,landethomas1944} can also be interpreted as dealing with a massive photon \cite{gratusperlicktucker2015}.

We spell out the Lagrangian density in SI base units. In the 4-vector potential $A^\alpha = (\phi/c,\vec{A})$, to the scalar component $\phi$, measured in Volt, correspond [kg m$^2$ s$^{-3}$ A$^{-1}$], to the vectorial component $\vec{A}$, measured in Volt/c, [kg m s$^{-2}$ A$^{-1}$]; to the electromagnetic field tensor $F_{\alpha\beta} = \partial_\alpha A_\beta - \partial_\beta A_\alpha$, measured in Volt/metre, correspond [kg m s$^{-3}$ A$^{-1}$]; to the permeability $\mu$, measured in Henry/metre, [kg m s$^{-2}$ A$^{-2}$] (in vacuum $\mu_0$ = 1.26 H m$^{-1}$); in the 4-vector current $j\!~^\beta = \left(\rho c, \vec j\right )$, to the charge density $\rho$, measured in Coulomb/metre$^3$, correspond [m$^{-3}$ s  A], to the current density, measured already in base units Amp\`ere/metre$^2$ [m$^{-2}$ A]. 
The additional dBP term contains ${\cal M} = m_\gamma c/{\hbar} = 1/ \lambdabar$ [m$^{-1}$], where $m_\gamma$ is the photon mass [kg],  
$\hbar = 1.05 \times 10^{-34}$ Joule$\cdot$second, that is [kg m$^{2}$ s$^{-1}$], and $\lambdabar$ the reduced Compton wavelength. Finally, the  speed of light $c = 2.99\times 10^8$ m~s$^{-1}$ is meant as the limiting - Maxwellian - velocity of the theory., The dBP Lagrangian density $\cal L$ reads, {\it e.g.}, \cite{greinerreinhardt1996,tulugi05,acciolynetoscatena2010a,goni10,spqigiro11}, turned into energy density SI units

\begin{equation}
{\cal L} = - \frac{1}{4\mu }F_{\alpha\beta}F^{\alpha\beta} + {j^\alpha} A_\alpha 
- \frac{{\cal M}^2}{2\mu}A_\alpha A^\alpha~.
\label{dBPLagr}
\end{equation}

Of the four Maxwell's equations, the dBP formalism modifies two by letting appear a mass dependent term: the divergence of the electric field (Coulomb-Gauss' law), and the curl of the magnetic field (Amp\`ere-Maxwell's law). 

The test of the former in a ground laboratory determined the mass upper limit of $2\times 10^{-50}$~kg \cite{wifahi71}. 
For the latter, analysis in the solar wind and Parker modelling led first to $m_\gamma < 10^{-52}$~kg at 1 AU \cite{ry97}, and later $m_\gamma < 1.5\times 10^{-54}$~kg at $40$ AU \cite{ry07}, limit accepted by the Particle Data Group \cite{oliveetal2014}. 
Nevertheless, such solar wind limits are far from being unquestionable \cite{retinospalliccivaivads2016}.   

A supposedly more stringent limit has been achieved through modelling of the hydromagnetic waves in the Crab Nebula \cite{basc75}, namely $3 \times 10^{-56}$ kg; therein the authors state that their arguments are not rigorous and open to doubt. A re-analysis has led to a mass upper value of $3 \times 10^{-63}$ kg \cite{ch76}. 
The difference of seven orders of magnitude for the same analysis is indicative of the scarce reliability of these approaches.      
Lower limits, as $10^{-62}$ kg,  have been claimed also when modelling the galactic magnetic field \cite{ya59,addvgr07}. 

The lowest theoretical limit on the measurement of {\it any} mass is prescribed by the Heisenberg uncertainty principle $m \geq \hbar/2 \Delta t~c^2$, and gives $1.2\times 10^{-69}$~kg, where $\Delta t$ is the currently supposed age of the Universe ($1.37 \times 10^{10}$ years). { Photon mass upper limits are not quantum mechanics direct limits, but indirect ones. They are deduced from the modifications to classic laws that an ensemble of massive photons would produce}.  

The examination of the literature inspires a critical attitude and prompt to question whether the limits are nothing more than the outcome of idealised models: "Quoted photon-mass limits have at times been overly optimistic in the strengths of their  characterizations. This is perhaps due to the temptation to assert too
strongly something one �knows� to be true" \cite{goni10}. We share this concern also for the actual solar wind limits, and feel safer 
if values larger than $m_\gamma = 10^{-54}$ kg are still considered partially under investigation.

For $m_\gamma = 0$, Eq. (\ref{dBPLagr}) reduces to the usual Maxwell Lagrangian. The Euler-Lagrange equation leads to the dBP wave equation

\beq
\left[\square + {\cal M}^2\right]A^\alpha = \mu_0 j^\alpha~,
\label{dBPwave}
\eeq
where the permeability $\mu_0 = 1.23 \times 10^{-6}$ H m$^{-1}$. 
Equation (\ref{dBPwave}) also holds for Stueckelberg's theory \cite{st57}, where the auxiliary (real) scalar field is a ghost in the sense that it  does not interact with any of the observable particles.  Its purpose is to make the theory manifestly gauge invariant.

For $m_\gamma\neq 0$, the speed of propagation \footnote{Photons can indeed be interpreted as quanta of harmonic waves. However, it must be noted that the concept of photon has its own independence. The nature of light is considered to be double, according
to the scale, with respect to the wavelength, at which we are observing it. Not only wave packets are composed of photons, but also plane waves. Further, if we refer to a quantum theory, its Hamiltonian will be given in terms of annihilation and creation operators. This means that photons with a defined wave numbers are destroyed or destroyed. These are supposed to be diffused wavefunctions, which, in the classical limit that is of interest here translate into to plane wave with a well defined wavelength. For such a wave, the local effects are negligible.} $v_g$ depends upon the frequency $f$ and is smaller than the Maxwellian speed of light $c$

\beq
v_g = c \left [1 - \left (\frac{c{\cal M}}{2\pi f}\right)^2\right ]^{1/2}~.
\label{vg}
\eeq 

At sufficiently high frequencies, for which the photon rest energy is small with respect to the total energy (in practice $f\gg 1$ Hz), the positive difference in velocity for two different frequencies ($f_2 > f_1$) is at first order \cite{db40}

\beq
\Delta v_g = v_{{\rm g}2} - v_{{\rm g}1} = \frac{c^3{\cal M}^2}{8\pi^2} \left(\frac{1}{f_1^2} - \frac{1}{f_2^2}\right)~,
\label{deltavg}
\eeq
being $v_g$ the group velocity. For a single source at distance d, the difference in the time of arrival of the two photons is

\beq
\Delta t = \frac{d}{v_{{\rm g}1}} - \frac{d}{v_{{\rm g}2}} \simeq \frac{\Delta v_g d}{c^2} = \frac{d {c\cal M}^2}{8\pi^2 }
\left(\frac{1}{f_1^2} - \frac{1}{f_2^2}\right) = \frac{d c^3m_\gamma^2}{8 \hbar^2\pi^2 }
\left(\frac{1}{f_1^2} - \frac{1}{f_2^2}\right)~. 
\label{deltatnu}
\eeq

Inserting all values in SI units, we get  
\beq
\Delta t~[{\rm s}]  \simeq  \frac{d~[{\rm m}]}{c~[{\rm ms}^{-1}]} 
\left(\frac{1}{f_1[{\rm s}^{-1}]^2} - \frac{1}{f_2[{\rm s}^{-1}]^2}\right) 
\left (10^{100}~{\rm s}^{-2} {\rm kg}^{-2}\right )
m_\gamma [{\rm kg}]^2~.
\label{deltatnusi}
\eeq

Equations (\ref{deltatnu},\ref{deltatnusi}) imply that low frequency photons arrive later than high frequency photons. In Eq. (\ref{deltatnusi}), we have multiplied and divided by c 
to get a mnemonic expression.  
Modern approaches to gravity quantisation, {\it e.g.}, \cite {amelino-cameliaetal1998,amelinocamelia2002,kowalskiglikman-nowak2002,magueijo-smolin2003,abdoetal2009,amelino-camelia2009,mavromatos2010,
bolmontjacholkowska2011,amelinocamelia2013,ellismavromatos2013}, lead also to variation of light speed versus the wavelength. Conversely to Eq. (\ref{deltatnu}), they manifest themselves at high frequencies, thereby the interest for Gamma-Ray Bursts.


\section{Pulsars, magnetars, fast radio bursts and dispersion}

\subsection{Pulsars and magnetars}

Pulsar and magnetars are neutron stars characterised by a high magnetic field $10^4-10^{11}$ T, fast rotation $10^{-1}-10^3$ Hz, and the  
emission of a beam of electromagnetic radiation, from radio to gamma rays, including bursts \cite{coti-zelatietal2015}. Few magnetars, though, are known to be active in the radio spectrum and above, such as the magnetar in the Galactic centre \cite{kaspietal2014,torneetal2015,pennuccietal2015}. While the rotation slow down of pulsars is used for timing, the variability of the rotation of magnetars induces a larger timing noise. 

The pulsar radiation passes through the interstellar medium (ISM), which temperature of ranges from $10$ to $10^7$ K. Free electrons in the ISM affect the radiation. Due to the dispersive nature of plasma, or the ionised part of the atoms, lower frequency radio waves travel through the ISM slower than higher frequency ones \footnote{For a more detailed
description of dispersion and waves in plasmas, see \cite{bekefi1966,melrose1980,nicholson1983,bastian2005}. Nevertheless, the literature on massive photons, {\it e.g.}, \cite{feinberg1969,goni71,tulugi05,tuyelu05} deals with plasmas as we deal with in this paper. Feinberg \cite{feinberg1969} pointed out that pulse
arrival times show no sign of any dispersion, except that
implied by the simple quadratic,
over the whole range of frequency from radio to optical. We invite observers to analyse any departure from this law.}.
The delay in the arrival of the pulses is directly measurable as the dispersion measure (DM) of the pulsar. The DM is the total column density of free electrons $n$ between the observer and the pulsar, and it is used to construct models of the free electron distribution. It is computed through the integral along the propagation path from the source to the observer 

\beq
DM = \int n dl~.
\eeq

Purging the raw measurements of the flux vis {\' a} vis the frequency through the DM is the first step in signal processing. 
The DM conveys valuable information about the location of a
burst or pulsed source. It gives an estimate of the path length and hence the source distance.
In pulsar timing, different arrival times of the incoming photons are routinely measured, but lacking any other independent measurement of the electron density of the ISM, the delays are solely attributed to plasma, or the ionised part of the atoms.  
The group velocity is given by \cite{lorimerkramer2005}

\beq
v_g = c \left [1 - \left (\frac{f_P}{f}\right)^2\right ]^{1/2}~,
\label{vgplasma}
\eeq 
being $f_{\rm P} = e/(2\pi) [n/(\epsilon_0 m_{\rm e})]^{1/2}$ the plasma frequency and $n$ the average electron density along the line of sight. 

Indeed, Eq. (\ref{vgplasma}) resembles Eq. (\ref{vg}). Again at first order, we have that the delay in pulse arrival times across a finite bandwidth due to the frequency dependence of the group velocity of the pulse on the ionised components of the ISM, is given by

\beq
\Delta t = \frac{d}{c}\frac{f_{\rm P}^2 }{2} 
\left(\frac{1}{f_1^2} - \frac{1}{f_2^2}\right)~.
\label{deltatnuplasma}
\eeq

Inserting all values in SI units, we get

\beq
\Delta t~[{\rm s}]  \simeq  \frac{d~[{\rm m}]}{c~[{\rm ms}^{-1}]} 
\left(\frac{1}{f_1[{\rm s}^{-1}]^2} - \frac{1}{f_2[{\rm s}^{-1}]^2}\right) 
\left ( 4.03 \times 10^{1} {\rm m}^3 {\rm s}^{-2}\right ) n~[{\rm m}^{-3}]~.
\label{deltatnuplasmasi}
\eeq

Incidentally using the Dispersion Measure (DM) in units pc cm$^{-3}$, that is $3.0857 \times 10^{22}$ m$^{-2}$, Eq. (\ref{deltatnuplasmasi})  may be cast as

\beq
\Delta t~[{\rm s}] \simeq 
\left(\frac{1}{f_1[{\rm s}^{-1}]^2} - \frac{1}{f_2[{\rm s}^{-1}]^2}\right) 
\left ( 4.17 \times 10^{15} {\rm pc}^{-1} {\rm cm}^3 {\rm s}^{-1}\right)~{\rm DM}~[{\rm pc}~{\rm cm}^{-3}]~.
\label{deltatnuplasmadm}
\eeq

The numerical values in Eq. (\ref{deltatnuplasmadm}) are arranged to let the reader use DM in pc cm$^{-3}$ units. 
Comparing Eqs. (\ref{deltatnusi},\ref{deltatnuplasmasi}), 
we get 

\beq
\frac {m_\gamma}{\sqrt {n}} \left[{\rm kg}~{\rm m}^{3/2} \right]= { 6.62 \times 10^{-50}}~,
\label{equivalence}
\eeq

which implies that for this ratio - in different units in \cite{feinberg1969,goni71,tulugi05,tuyelu05} - a massive photon and the average electron density along the line of sight determine the same dispersion. 

Anisotropy, inhomogeneity, and time variability of the electron density (and thereby turbulence, scintillations, multipaths) are the main constraints in establishing an upper limit to the photon mass. 
 
Pulsar observations 
\cite{bawh72} indicated that the speed of light was constant to within $10^{-20}$ throughout the visible, near
infrared and ultraviolet regions of the spectrum, corresponding to a rough upper limit to the
photon mass of $3 \times 10^{-49}$ kg \cite{bawh72,tuyelu05,acciolynetoscatena2010a}.  
It is evident that exploring lower frequencies provides more stringent upper limits to the photon mass. 

We remark in passing that also millisecond pulsars \cite{schaefer1999,tuyelu05} have been considered in the context of 
testing photon mass.

\subsubsection{Superdispersion}
\paragraph{Pulsars}

The LOw Frequency ARray (LOFAR) \cite{vanhaarlemetal2013} is now operating and getting data at 15 MHz. LOFAR, Westerbork and Lovell Telescope frequencies $15-1400$ MHz, are used \cite{piliaetal2016} also to analyse superdispersion, {\it i.e.} an excess of dispersion at lower frequencies. 

Superdispersion has been detected by some  
\cite{shitovmalofeev1985,kuzmin1986,shitovmalofeevizvekova1988,hankinsetal1991,pennuccietal2015,cordesshannonstinebring2015}, while others have not found evidence \cite{hassalletal2012,piliaetal2016}.
In \cite{phillipswolcszan1992,ahujaguptamitrakembhavi2005,ahujamitragupta2007} the larger dispersion is at higher frequencies, but for a limited number of sampling frequencies. 

One possible explanation of superdispersion is based on the inhomogeneous properties of the ISM that, by  
causing multi-path propagation of radio waves where the path depends on the frequency, determines the sampled column density of
free electrons being also a function of frequency. Thereby, the DM inconsistencies might be the consequence of imposing a 
$1/f^2$ simple law onto a frequency-dependent and more complex dispersion due to an inhomogeneous ISM \cite{pennuccietal2015}.  

\paragraph{Magnetars}

In \cite{pennuccietal2015}, for the magnetar SGR J1745-2900, the authors find superdispersion between the L and S bands. They exclude causes as low quality data, profile misalignments, modelling and systematic errors, time variability. 
On the other hand, the authors observe that frequency dependencies other than $1/f^2$ appear to produce DM inconsistencies.
We complement their conclusions by commenting that it is not evident to assume that the medium inhomogeneities, and thereby the multipaths, determine always the same behaviour versus frequency. Whether, this is due to an extra-dispersion $1/f^2$ caused by a massive photon cannot be claimed, even theoretically at this stage.  

Indeed, for magnetars we are compelled to consider the presence of the overcritical magnetic field. The latter demands the analysis of alternative electromagnetism \cite{heeu36} to see if delays can be caused by the large magnetic field. In \cite{bopbsp2016}, there is the description of how photons red or blue shift exiting the magnetar and travelling towards the observer at infinity.

\subsection{Fast radio bursts}

Fast Radio Bursts (FRB) are millisecond bursts of radio radiation, not yet consensually identified with an astronomical object or phenomenon, and discovered in pulsar surveys. The bursts are strongly dispersed, possibly hinting that FRBs are at cosmological distances with redshifts in the range 0.3-1.3 \cite{dennison2014,tuntsov2014,keaneetal2016,katz2016}. 

They are characterised by the fluence (dimensions of action per surface area), time integral of the flux (dimension of energy per surface area) 

\beq
{\cal F} = \int F (t) dt~. 
\eeq
   
The fluence is measured in Jy ms (Jansky millisecond, where a Jansky is 10$^{-26}$ W m$^{-2}$ Hz $^{-1}$). Seventeen FRBs have been catalogued so far \cite{petroffetal2016}.    

FRB have large values of DM, though they are observed at high galactic latitude. After subtraction of the (small) estimated Galactic component, the remaining DM must have another origin. Leaving aside massive photons, there are two conventional explanations. One is the close plasma environment of the galactic source, the other is the intergalactic plasma. 

FRBs have been used recently to set photon mass upper limits \cite{boelmasasgsp2016,wuetal2016b} in the GHz region. Meanwhile, FRB are also  targeted \cite{trotttingaywayth2013,karastergiouetal2015,obrockastapperswilkinson2015,thyagarajanbeardsleybowmanmorales2015,tingayetal2015,
rowlinsonetal2016} at lower frequencies by numerous radio telescopes like LOFAR \cite{vanhaarlemetal2013}, the Long Wavelength Array (LWA)  \cite{ellingsonetal2013}, and the Murchison Widefield Array (MWA) \cite{bowmanetal2013}.

In the future, the different redshift dependences of the plasma and photon mass contributions to DM can be used to improve the sensitivity to the photon mass if more FRB redshifts are measured. For a fixed fractional uncertainty in the extra-galactic contribution to the DM of an FRB, one with a lower redshift would provide greater sensitivity to the photon mass \cite{boelmasasgsp2016}.

\section{Discussion}
\label{discussions}

This research note addresses massive electromagnetism \footnote{Instead, we have not addressed nor the classic or quantum frequency shifts \cite{ginzburg1961,wolffoley1989,foleywolf1989,laiorizzitartaglia1997} nor Doppler shifts nor the cosmological redshift.}. 
We have four areas of investigations in front of us. We spell them out in form of questions 
\begin{itemize}
  \item{Is usual pulsar dispersion hiding a massive photon?. There is abundant literature showing that the upper limits to the photon mass from traditional analysis of pulsars are not competitive with other estimates. The latter, though, often are product of theoretical models, and not real experiments.}
  \item{Is pulsar superdispersion (also) a manifestation of a massive photon? This is a new line of investigation, but we feel that the question will be unanswered as long as a mathematical model of superdispersion does not exist.}
  \item{Is magnetar superdispersion (also) a manifestation of a massive photon? The results from the study \cite{pennuccietal2015}, for the magnetar SGR J1745-2900 need to be confirmed. Further, the effects of the overcritical magnetic field must be separated out. }   
  \item{Is the extra dispersion from FRBs (also) a manifestation of a massive photon? This question will be unanswered as long as the distance of the FRB sources remains uncertain.}     
\end{itemize}

Disentangling the distance to sources objects from dispersion appears one main obstacle. Lacking the ability of estimating distances
independently from dispersion, the similarity of massive photon and electron density dispersions obliges to set only upper limit to the photon mass. Possibly, double pulsars \cite{krameretal2006} might provide a mean to disentangle distance from both dispersions, but unlikely to assign the nature of such dispersion. 

Another hope for disentanglement might be to study signals around the cusp of the plasma frequency $f_P$. Around and below such frequency, 
the group velocity would differ from the $1/f^2$ behaviour, while massive photon dispersion would keep its mathematical profile. 

When referring also to gravitational wave detection studies with pulsar timing arrays, dispersion emerges as a crucial issue. 
Time variable delays due to radio wave propagation in the ionised interstellar medium are a substantial source of error 
\cite{Stinebring2013,LamCordesChatterjeeDolch2015,Levin2015,Pallyaguruetal2015}. We further remark that setting upper limits to the photon mass may be mandatory to set limits to the graviton mass, or testing general relativity and the equivalence principle, when the two travel speeds are compared, {\it e.g.}, \cite{cutlerhiscocklarson2003,deffayetmenou2007,kocksishaimanmenou2008,NishizawaNakamura2014,branchinadedomenico2016,liuzhaoyouluxu2016,wuetal2016a}. While for general relativity, the graviton mass is null, it is not the case for the numerous alternative theories
\cite{goni10,will14} \footnote{ In \cite{abbottetal2016}, there is an upper limit to the graviton mass of $2\times 10^{-58}$~kg, based on a similar expressions to Eq. (\ref{vg}), going as $1/f^2$. For clarity, this upper limit does not rely on a comparison with photon times of arrival but solely on the gravitational waveforms. Obviously, such limit is not to be intended as on a single graviton, but on the dispersion that an ensemble of massive gravitons would produce macroscopically. Otherwise, for the Heisenberg principle the strict measurement of the mass upper limit as $2\times 10^{-58}$~kg on a single graviton would demand an observation time of approximately a month. For a review on photon and graviton mass upper limits, see \cite{goni10}.}.

What studies should target? 1) an analysis of the frequency scaling as $af^{-1} + bf^{-2} + cf^{-3} + df^{-4}$; 2) an improvement of data quality in terms of timing at low frequencies; 3) an analysis of time, space, and frequency variability of the electron density; 4) extensive studies at very low frequencies, possibly of sources at large distances, towards sky  directions where electron density dispersion is supposedly not large. 

A contribution to the answers to these queries might be provided by opening a new window in the sub-MHz region.      

\section{OLFAR}

Opening a new window in the sub-MHz region consists of a few challenges. First of all the wavelength of the signals is in the order of tens of meters, meaning that usually large antennas are needed (or
synthetic apertures in the case of array antennas). Usually this is done by creating large arrays
instead of single dish antennas. Secondly for very low frequencies, the Earth's ionosphere blocks
frequencies below 30 MHz. 
To overcome this problem, initiatives based on satellite systems have been proposed. Because
of the large baselines needed for low frequencies, array antennas are the only feasible option
in space. One of the first initiatives is the DARIS project~\cite{rajanetal2016}. DARIS is a scenario with eight slave spacecraft and a central spacecraft, in which the nodes will do the sensing part while the mothership will have additional processing and communication tasks. One of the conclusions of the study was that technology has reached a level where this type of scenario is realistic and can be
implemented with commercial off-the-shelf components. However, having a central spacecraft will increase the risk of failure of the system.

The following up project for radioastronomy is the Orbiting Low Frequency Antennas for Radio astronomy (OLFAR) \cite{verhoevenetal2011,deckensetal2014,budianuetal2015,rajanetal2016} intends to map the celestial sources in the 0.1-10 MHz range and be a tool for navigation \cite{tartagliaetal2011}.
OLFAR plans to place a swarm of satellites into a Moon orbit. The purpose of these
satellites is to function as a distributed low frequency array far away from man-made Radio
Frequency Interference (RFI) and away from the blocking ionosphere.
A swarm of 50 or more nano-satellites orbiting faraway from terrestrial interference will be used to sense and sample the cosmic signals, process the information by means of distributed correlation,
and send the processed data to a base station on Earth. Thus, each member of the swarm will have to
fulfil three main tasks: radio observation, data distribution and processing, and down-linking. From the hardware point of view separate subsystems will have to be designed and integrated on a miniaturised spacecraft platform. Multiple antenna systems will have to be hosted by the nano-satellites
to be able to support the data flow into, within, and out of the swarm.

To prevent terrestrial radio frequency interference (RFI), a Moon orbit or the Earth-Moon Lagrangian point behind the Moon (L2 point) is considered. When the swarm is  behind the Moon, the latter can act as a shield against the RFI produced by the Earth. This will reduce the amount of RFI coming from Earth. However,  due to diffraction, the RFI might bend around the Moon, and still have significant presence at locations where there is no direct line of sight to Earth, especially at the lowest frequencies. The amount of RFI presence behind the Moon due to diffraction is not known, but can be predicted. 

Interference produced at the surface of the Earth, artificial radio transmissions and lighting, are
effectively blocked by the Moon. Below 3 MHz, these signals are already attenuated a lot by the
ionosphere of the Earth, and above 3 MHz the attenuation during ground wave propagation around
the Moon is so high that the surface produced RFI will not be observed behind the Moon.
The Auroral Kilometric Radiation (AKR) is more troublesome because it is not produced at the
surface, but at an altitude of 1-3 times the radius of the Earth. This means that it does not have to
propagate through the ionosphere to reach the Moon. The AKR has a frequency range of 100 kHz to
1 MHz. These low frequencies are less attenuated by ground wave propagation. However, due to the
very low conductivity of the Moon surface and the very rough terrain, the RFI produced by AKR will
still be much weaker than the Galactic background radio noise when there is no direct Line Of Sight (LOS).
However, due to the high altitude of the AKR, there will be a smaller area behind the Moon where
there is no direct LOS to the source, which reduces the amount of time where observations can be
executed during a Moon orbit. For a satellite at the Earth-Moon L2 point it is even worse. This point
is located outside the LOS-free area, and therefore satellites located there will be directly exposed to
RFI produced by the AKR. Because the interference will be very significant, this location is not
suitable for OLFAR. Another issue is the reflection of the AKR at an altitude of 20-40 times the radius of the Earth. During these reflections that occur approximately 10\% of the time, RFI produced by AKR will directly travel to the backside of the Moon and OLFAR will not be able to do observations in orbits higher than approximately 1000 km. During the other 90\% of the time, observations will be possible.

\subsection{OLFAR contribution to fundamental physics}

OLFAR will open the unexplored frequency range of 0.1-10 MHz. By doing so, new astrophysical sources or new signal from know sources 
will emerge \cite{lacki2010}. Meanwhile, it will contribute to answer to the queries posed in Section~\ref{discussions}. 

\begin{figure}
\centering \includegraphics[width=14cm]{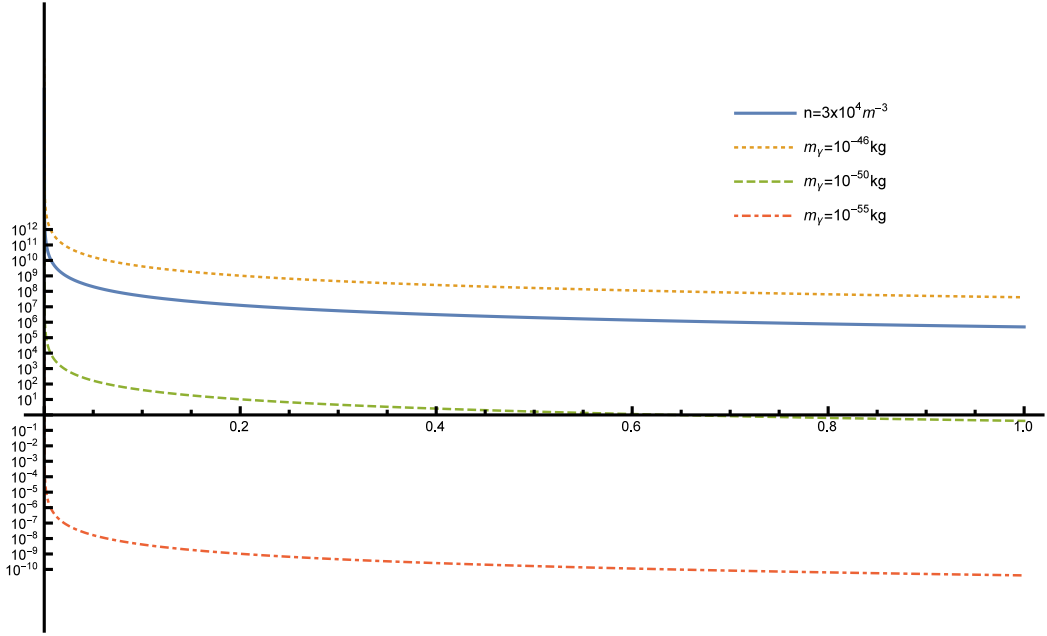}
\caption{{ For illustration, electron density and dBP dispersions are plotted versus the horizontal and vertical axis, indicating the frequency [MHz] and the delay [s], respectively. Equations (\ref{deltatnu},\ref{deltatnuplasma}) are applied to a distance $d = 4$ kpc (representative pulsar distance), an average electron density along the line of sight $n=3\times 10^4$ m$^{-3}$ (thick blue line), photon mass $m_\gamma = 10^{-46}$ kg (dotted orange line), $10^{-50}$ kg (dashed green line) and $10^{-55}$ kg (dotted-dashed red line). The frequency range is $0.1-1$ MHz and the delays are computed with respect to the time of arrival
at high frequencies. For these values, the dispersion of the massive photon is either larger or smaller than the contribution of the average electron density along the line of sight. The dispersion value at $1$ MHz for $m_\gamma = 10^{-46}$ kg is $4.11 \times 10^7$ s 
(around 1.3 years), for $10^{-50}$ kg is 0.411 s, for $10^{-55}$ kg is 4.11 $\times 10^{-11}$ s. The dispersion for the electron density is $4.96 \times 10^5 $ s (about 6 days). }}
\label{fig1n}
\end{figure}

Figure \ref{fig1n} shows four dispersion curves due to photon mass and the average electron density along the line of sight in frequency band of OLFAR.

Applying Eqs. (\ref{deltatnu}-\ref{equivalence}) at $1$ MHz, we get for a typical distance $d = 4$ kpc (representative pulsar distance), the values in Tab. \ref{tab1}.  
{ They show that a given delay might be attributed to the electron density in the line of sight or else to a supposed massive photon, Eq. \ref{equivalence}. More subtly, a delay discrepancy at low frequencies, may hide something more fundamental than the variation of dispersion. }

{\renewcommand{\arraystretch}{1.2}
\begin{table}[h]
\centering
\caption{{ The equivalent theoretical delays at $1$ MHz due to the electron content or massive photon dispersions, Eqs. (\ref{deltatnu}-\ref{equivalence}), and for a typical distance $d = 4$ kpc (representative pulsar distance), computed with respect to the time of arrival
at high frequencies.}}
\vskip 10pt
\label{tab1}
{ 
\begin{tabular}{|l|c|c|}
\hline
{\bf $\Delta t$ [s]}                                
& {\bf $m_\gamma$ [kg]}                        
& {\bf $n$ [m$^{-3}$]}                    
\\ \hline
$10^{-10}$
& $1.6 \times 10^{-55}$                           
& $6 \times 10^{-12}$\\ \hline
$10^{-8}$
& $1.6 \times 10^{-54}$                           
& $6 \times 10^{-10}$\\ \hline
$10^{-6}$
& $1.6 \times 10^{-53}$                           
& $6 \times 10^{-8}$\\ \hline
$10^{-4}$
& $1.6 \times 10^{-52}$                           
& $6 \times 10^{-6}$\\ \hline
$10^{-2}$
& $1.6 \times 10^{-51}$                           
& $6  \times 10^{-4}$\\ \hline
$1$
& $1.6 \times 10^{-50}$                           
& $6    \times 10^{-2}$\\ \hline
$10^{2}$
& $1.6 \times 10^{-49}$                           
& $6                 $\\ \hline
$10^{4}$
& $1.6   \times 10^{-48}$                           
& $6 \times 10^{2}$\\ \hline
$10^{6}$
& $1.6 \times 10^{-47}$                           
& $6   \times 10^{4}$\\ \hline
$10^{8}$
& $1.6 \times 10^{-46}$                           
& $6   \times 10^{6}$\\ \hline
\end{tabular}
}
\label{orleans}
\end{table}
}

\subsection{Expected signals and processing} 

The availability of this extended frequency range makes it possible to study pulsars and pulsar timing in a frequency range in which most pulsars are intrinsically brightest. 
However, despite of this, the vast majority of pulsars have been discovered and studied at higher frequencies, in the range 300-2000 MHz. There are three main reasons for this: the deleterious effects of the ISM on pulsed signals; the effective background sky temperature of the Galactic synchrotron emission; and ionospheric effects. All of the effects become worse towards lower frequencies. With OLFAR the ionospheric effects are eliminated, leaving the ISM effects and the high background sky temperature. Measuring dispersion measure variations and modelling the scattering of pulse profiles at the OLFAR frequencies has the potential to help improve pulse shapes and thus enhance high precision timing. At the same time, this means we can study the ISM.

Timing in the OLFAR swarm has been studied as part of the synchronisation for the swarm 
\cite{rajanbentumboonstra2013}. The requirements of the clock in each of the satellites can be briefly summarised as:
\begin{enumerate}
\item Sampling jitter  $\Delta t_{jitter}$
\begin{enumerate}
\item[(a)] $\Delta t_{jitter} < 10$ ps for 8 bit sampling
\item[(b)] $\Delta t_{jitter} < 1$ ps for 12 bit sampling
\end{enumerate}
\item Allan deviation $\sigma_\zeta (\tau_c)$
\begin{enumerate}
\item[(a)] Short term $\sigma_\zeta (\tau_c) \leq 10^{-8}$ for $\tau_c$ = 1 second
\item[(b)] Long term  $\sigma_\zeta (\tau_c)  \leq 10^{-11}$ for $\tau_c$ = 1000 seconds
\end{enumerate}
\end{enumerate}

To achieve Allan deviations of order $10^{-8}$ and $10^{-11}$ the usual solutions are Rubidium standards and Oven controlled Crystal oscillators (OCXO). Although cesium and maser families can offer orders of magnitude lower Allan deviations, they are also very expensive, in terms of mass and power for an OLFAR satellite and hence not considered. There has been consistent research in developing chip scale atomic clocks, especially on rubidium clocks, \cite{knappeetal2006}) based on
Vertical Cavity Surface Emitting Lasers (VCSELs), which enable orders of magnitude reduction in size and power. The Quantum SA.45s is a Rubidium CSAC, Chip Scale Atomic Clock, which is based on VCSEL and meets the Allan deviation requirements of OLFAR up to 1000 seconds. This CSAC weighs less than 35 grams and has a steady power consumption of 125 mW. This  suits the requirements on an OLFAR node.

\subsection{Other opportunities}

Ground observatories at low radio frequencies include the already mentioned LOFAR \cite{vanhaarlemetal2013} and NenuFAR \cite{zarkaetal2016}. 

Naturally, we recall the potential offered by the Square Kilometre Array (SKA) \cite{lazio2013,salzanodabrowskilazkoz2016} operating at ns level.  

Supposing an ambitious sub-picosecond precision and ideal conditions, a low Earth Orbit satellite as the Atomic Clocks Ensemble  in Space (ACES) \cite{lounisetal1993,aces1994} would determine an $m_\gamma$ upper limit around $10^{-46}$ kg; instead the Space-Time Explorer and Quantum Equivalence Principle Space Test (STE-QUEST) \cite{altschuletal2015} could use the highly elliptic orbit to get around $10^{-47}$ kg; a large space mission as LISA \cite{decheretal1980,bertotti1984} with million kilometer interferometric arms around $10^{-48}$ kg; finally, a mission to the frontiers of the Solar System as Neptun, Pluto and the Kuiper belt around around $10^{-51}$ kg. All the estimates refer to the S band.

\section{Conclusions and perspectives}

We have focused on the interest raised by the newly operating and future ground and space detectors at very low radio frequencies. The foundations of electromagnetism could be tested, confirming our beliefs or else contributing to establish new physics. 
Studies on dispersion appear crucial to unveil the causes of the excess of DM, and disentangle the measure of DM from distance. Low frequency observatories placed faraway from the Earth will be an essential aid to such studies, and will possibly set competitive limits to the photon mass. 

Meanwhile, theoretical investigations on the plausibility and implications of massive photons is to be pursued. 
In the context of Standard Model Extensions (SMEs), four general classes of 
Super Symmetry (SuSy) and Lorentz Symmetry (LoSy) breaking were analysed, leading to observable imprints at our energy scales \cite{bodshnsp2016a}. The photon dispersion relations show a non-Maxwellian behaviour for the CPT (Charge-Parity-Time reversal symmetry) odd and even sectors. The group velocities exhibit also a directional dependence with respect to the breaking background vector (odd CPT) or tensor (even CPT).  In the former sector, the group velocity may decay following an inverse squared frequency behaviour. Thus, a massive and gauge invariant Carroll-Field-Jackiw photon term in the Lagrangian has been extracted and the induced mass shown to be proportional to the breaking vector. The latter is estimated by ground measurements and leads to a photon mass upper limit of $10^{-19}$ eV or $2 \times 10^{-55}$ kg.

Implications for cosmology by non-linear and massive photon theories, and generally non-Maxwellian behaviour have not yet been evaluated adequately. Finally, laboratory experiments should be pursued, investigating in all directions, including the search of frequency shifts, incidentally using the same equipment to set upper limit to the Hubble parameter at small scale \cite{shamirfox1967,bonnor1999,dumin2012,priceromano2012,kopeikin2015}.     

 \section*{Acknowledgements} 

This work was partly carried during a visit of LB and ADAMS to the Centro Brasileiro de Pesquisas F{\'i}sicas  in Rio de Janeiro. The latter also acknowledges discussions with J.-C. Pecker (Coll\`ege de France), C. Salomon (Ecole Normale Sup\'erieure, Paris), A. Possenti and M. Pilia (Osservatorio Astronomico di Cagliari), L. Iess (Universit\`a di Roma, La Sapienza). 




\end{document}